\documentclass[letterpaper]{article} %
\usepackage{aaai2027}  %
\usepackage[hyphens]{url}  %
\usepackage{graphicx} %
\usepackage{natbib}  %
\usepackage{caption} %
\usepackage{algorithm}
\usepackage{algorithmic}

\usepackage{newfloat}
\usepackage{multirow}
\usepackage{pifont}
\usepackage{amssymb} %
\usepackage{amsmath}
\usepackage{amsmath}
\usepackage{listings}
\newcommand{\cmark}{\ding{51}}
\newcommand{\xmark}{\ding{55}}
\DeclareCaptionStyle{ruled}{labelfont=normalfont,labelsep=colon,strut=off} %
\floatstyle{ruled}
\newfloat{listing}{tb}{lst}{}
\floatname{listing}{Listing}

\usepackage{booktabs}
\nocopyright

\title{Audio-Visual World Models: Learning Physically Grounded Multisensory Dynamics}
\author{
    Jiahua Wang\textsuperscript{\rm 1}\equalcontrib, Leqi Zheng\textsuperscript{\rm 1}\equalcontrib, Jialong Wu\textsuperscript{\rm 1}, Yaoxin Mao\textsuperscript{\rm 2}, Shijie Cheng\textsuperscript{\rm 1}
}
\affiliations{
    \textsuperscript{\rm 1}Tsinghua Unniversity, 
    \textsuperscript{\rm 2}Beijing Institute of Technology
}

\begin{document}

\maketitle

\begin{abstract}
World models simulate environmental dynamics to enable embodied agents to plan and reason about future states. While real-world perception is inherently multimodal, existing approaches focus primarily on visual observations, leaving crucial spatial and temporal acoustic cues underexplored. In this work, we present a unified formulation of Audio-Visual World Models (AVWM), casting multimodal environment simulation under action control as a partially observable Markov decision process with synchronized audio-visual observations. As a foundational benchmark, we construct AVW-4k, comprising 30 hours of action-annotated binaural audio-visual trajectories across 76 indoor environments. To capture these physically grounded multisensory dynamics, we propose AV-CDiT (\textbf{A}udio-\textbf{V}isual \textbf{C}onditional \textbf{Di}ffusion \textbf{T}ransformer), featuring a novel modality expert architecture that balances visual and auditory learning, optimized via a three-stage training strategy. Extensive experiments demonstrate that AV-CDiT achieves high-fidelity prediction across both visual and auditory modalities. Furthermore, we validate its practical utility in embodied navigation, showing that AVWM significantly enhances a pretrained agent in continuous audio-visual navigation tasks.
\end{abstract}

\section{Introduction}
\label{sec:intro}

  Large-scale pre-trained models have substantially advanced a broad range of AI applications. A natural next step is to equip such models with an internal understanding of how the world evolves. World models~\cite{guo2025mineworld,zuo2025gaussianworld,hafner2025mastering,zhao2025drivedreamer,ding2025understanding} aim to encode the underlying dynamics of an environment, allowing an agent to predict future observations based on its actions, thereby providing a foundation for planning, decision making, and reasoning in complex environments.

  While most prior world modeling efforts focus on visual observations, real-world perception is inherently multimodal.
  Audio provides spatial and temporal cues that complement vision~\cite{du2025crab, ryu2025seeing, zhao2025audio,
  zhao2025audio2}, making it particularly valuable for embodied agents that must anticipate the multisensory
  consequences of their actions before acting. For such agents, a key requirement is \emph{physical consistency} across
  modalities: visual observations and acoustic signals should evolve coherently under agent actions, scene geometry, and
  sound propagation over time. This goes beyond perceptual plausibility or semantic correspondence, requiring coupled
  sensory predictions to remain physically grounded.

  Therefore, developing world models that jointly capture physically consistent audio-visual dynamics under low-level
  action control is an important step toward richer environmental understanding. However, existing research faces two
  major limitations that hinder progress toward multimodal world modeling.

\begin{figure}[t]
    \centering
    \includegraphics[width=0.8\columnwidth]{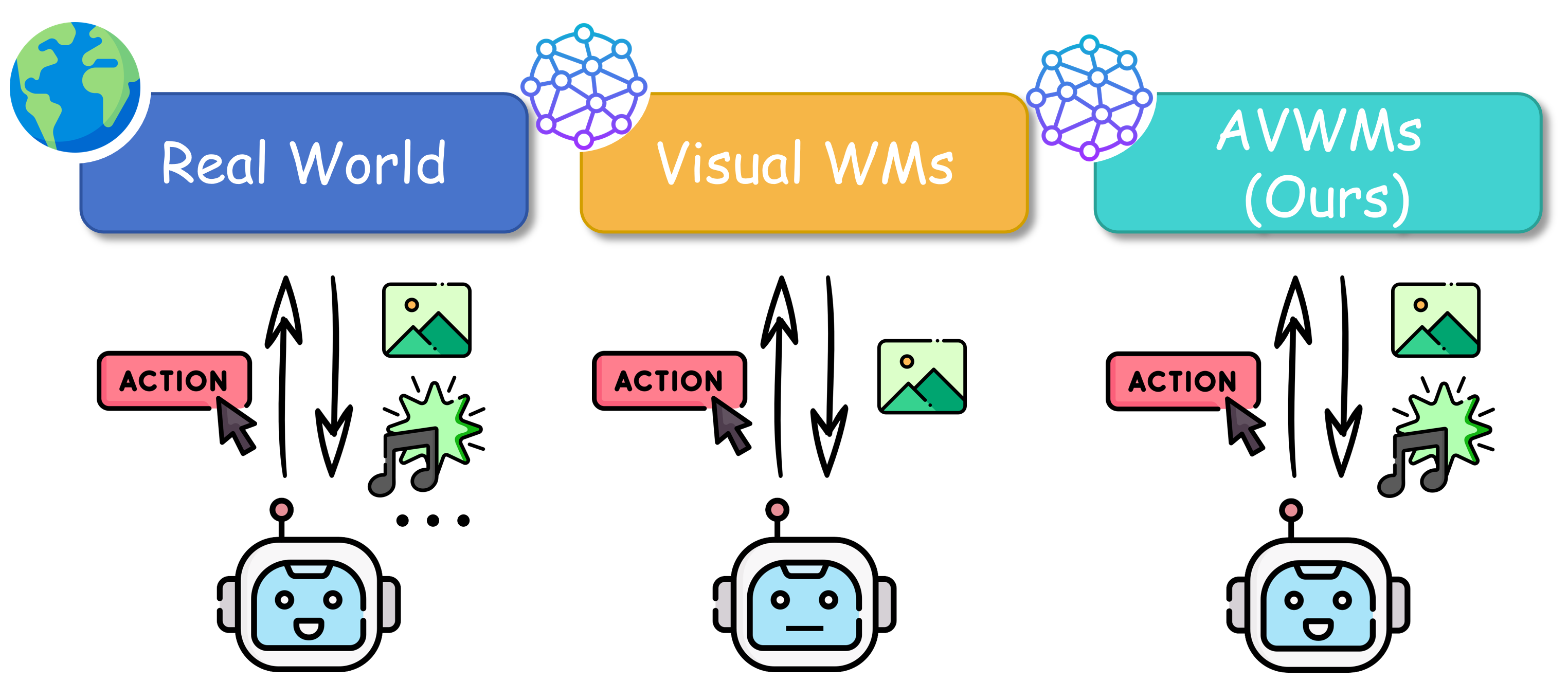}
    \caption{From unimodal to audio-visual world models. Our work introduces AVWMs to jointly simulate physically grounded audio-visual dynamics.}
    \label{fig:teaser}
\end{figure}

 \begin{table}[t]
    \centering
    \small
    \setlength{\tabcolsep}{3pt}
    \begin{tabular}{@{}lccc@{}}
    \toprule
    \textbf{Method} & \textbf{Modality} &
    \shortstack[c]{\textbf{Low}\\\textbf{Act.}} &
    \shortstack[c]{\textbf{Bin.}\\\textbf{Aud.}}\\
    \midrule
    DIAMOND~\cite{alonso2024diffusion} & Vision & \cmark & \xmark \\
    Genie~\cite{bruce2024genie} & Vision & \cmark & \xmark \\
    Cosmos~\cite{agarwal2025cosmos} & Vision & \xmark & \xmark \\
    VLWM~\cite{chen2025planning} & Vision+Text & \xmark & \xmark \\
    RLVR-World~\cite{wu2025rlvr} & Vision+Text & \cmark & \xmark \\
    \midrule
    \textbf{AV-CDiT} & Vision+Audio & \cmark & \cmark \\
    \bottomrule
    \end{tabular}
    \caption{Comparison with existing world modeling approaches. \textit{Low Act.} and \textit{Bin. Aud.} denote low-level action and binaural audio, respectively.}
    \label{tab:comparison}
  \end{table}

  \textbf{Limitation 1: Lack of formulation and benchmarks.}
  As illustrated in Figure~\ref{fig:teaser}, despite the importance of multimodal perception for embodied intelligence, existing world models remain confined to unimodal visual prediction. There is no commonly adopted formulation of
  audio-visual world modeling.
  Prior work has not clearly specified how synchronized visual observations, binaural audio, and fine-grained actions
  should be jointly represented and predicted.
  Meanwhile, existing datasets are either visual-only or contain audio-video recordings without action-conditioned
  correspondence, and rarely preserve spatial acoustic properties needed to study sound changes under agent motion
  and scene geometry.

  \textbf{Limitation 2: Lack of architectures for controllable audio-visual dynamics.}
  Existing world models are not designed to jointly predict physically coupled visual and acoustic streams under low-
  level actions.
  As summarized in Table~\ref{tab:comparison}, most focus on visual dynamics, while multimodal extensions mainly capture
  semantic associations rather than temporally synchronized sensory evolution.
  They also lack simultaneous audio-visual generation conditioned on agent control, making them insufficient for
  modeling how sound and vision co-evolve over time.

This raises a critical question: \textit{How can we formulate and implement an audio-visual world model that captures physically grounded multisensory dynamics under low-level action control?}

To bridge these gaps, we re-examine world modeling from a multimodal perspective, arguing that a genuine world model must capture how actions propagate through both the visual and acoustic domains over time.
Because this problem space remains largely unexplored, we take a ground-up approach and first establish a controlled setting in which physically grounded action-conditioned audio-visual dynamics can be isolated, modeled, and evaluated clearly.
In summary, our contributions are threefold:

\textbf{(1) Problem Formulation and Dataset.}
To address the first limitation, we provide a unified POMDP-based formulation of Audio-Visual World Models (AVWM), offering a common interface for modeling spatial audio, visual observations and low-level embodied actions within a single action-conditioned prediction setting.
To support this formulation, we construct AVW-4k, a controlled benchmark comprising approximately 30 hours of synchronized binaural audio-visual data with action annotations across 76 indoor environments (Section \textit{Formulation and Dataset}).

\textbf{(2) Modality Expert with Stagewise Training.}
To address the second limitation, we design AV-CDiT, an Audio-Visual Conditional Diffusion Transformer featuring a novel modality expert architecture that balances cross-modal interaction while preserving modality-specific representations.
We further develop a three-stage training strategy that ensures stable optimization and enhances multimodal integration (Section \textit{Methodology}).

\textbf{(3) Comprehensive Evaluation and Embodied Planning.}
We conduct extensive experiments demonstrating that our model achieves strong multimodal prediction quality together with robust physical consistency, and serves as an effective planning tool for continuous audio-visual navigation tasks, validating its utility for downstream embodied AI applications (Section \textit{Experiments and Results}).

\section{Related Work}
\label{sec:related_works}

\noindent\textbf{Multimodal World Models.}
World models aim to simulate environment dynamics through internalized commonsense knowledge~\cite{wu2024ivideogpt}. Unimodal models such as DIAMOND~\cite{alonso2024diffusion}, Genie~\cite{bruce2024genie}, and Cosmos~\cite{agarwal2025cosmos} achieve strong visual prediction, but lack the multisensory feedback needed for embodied intelligence. Recent studies~\cite{wu2026visual} highlight complementary roles of different modalities; we argue that audio is particularly important because it provides temporal and non-line-of-sight cues unavailable from vision alone.

\noindent\textbf{Audio-Visual Generation and Simulation.}
Recent audio-visual generation models~\cite{ruan2023mm,sung2023sound,xing2024seeing,kim2024versatile,chen2026skyreels} produce perceptually plausible and semantically aligned outputs across modalities, but are not designed for action-conditioned future prediction. In parallel, simulators such as SoundSpaces~\cite{chen2020soundspaces,chen2022soundspaces} provide physically grounded visual rendering and acoustic propagation, yet rely on privileged scene information rather than learned generalization to unseen environments.

\section{Formulation and Dataset}
\label{sec:formulation}

\subsection{Audio-Visual World Models}

An Audio-Visual World Model (AVWM) is designed to simulate an external environment containing sound signals.
In this environment there exists a stationary sound source.
We model this environment as a partially observable Markov decision process (POMDP), defined as a tuple $(\mathcal{S}, \mathcal{O}, \mathcal{A}, p)$.
At each time step $t$, $s_t \in \mathcal{S}$ represents the underlying state of the environment.
$o_t = \phi(s_t) = \{o_t^v, o_t^a\} \in \mathcal{O}$ represents the partial observation received by the agent, only providing incomplete information of $s_t$.
$o_t^v \in \mathbb{R}^{H \times W \times 3}$ is the visual observation (an image frame) and $o_t^a \in \mathbb{R}^{L \times 2}$ is the binaural auditory observation (a short segment of binaural audio).
Action $a_t = (u_t, \omega_t) \in \mathcal{A}$ specifies the low-level egomotion of the agent at time step $t$,
where $u_t$ and $\omega_t$ denote the translation and heading-change components of that action, respectively.

Note that all actions in the discrete action space $\mathcal{A}$ are executed with the same time duration.
When $a_t$ is executed at time step $t$, the environment transitions according to $p(s_{t+1} \mid s_t, a_t)$, yielding new audio-visual observations.

\begin{figure*}[t]
  \centering
   \includegraphics[width=0.95\textwidth]{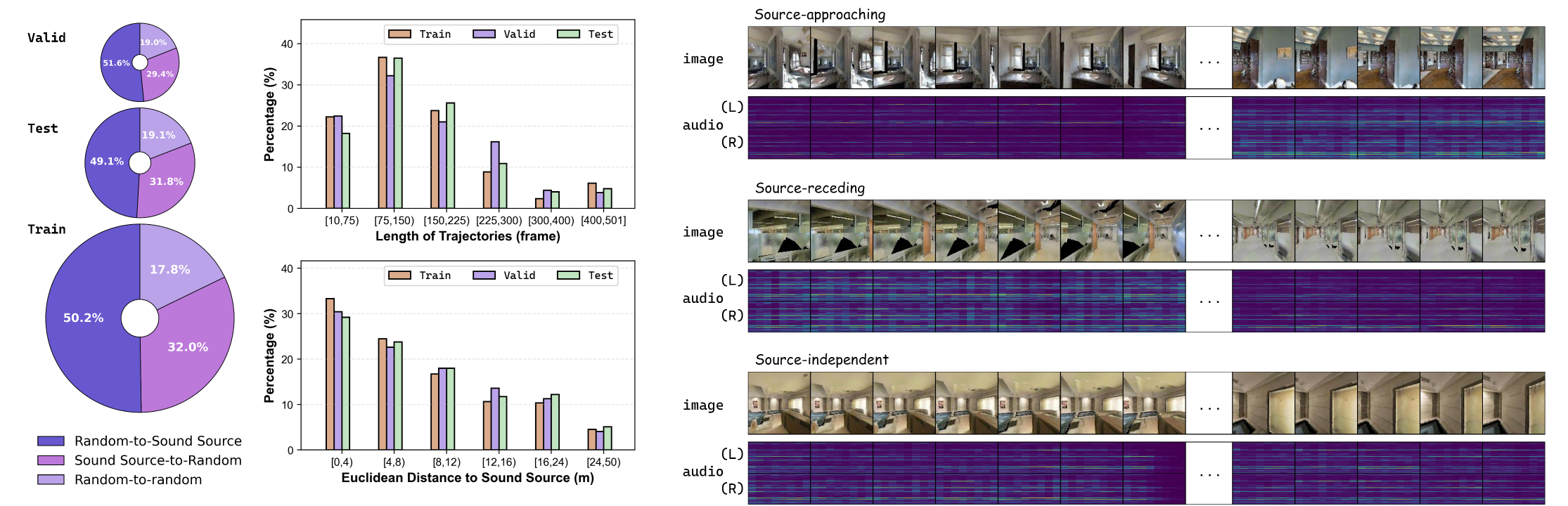}
   \caption{Dataset statistics and trajectory examples of the proposed AVW-4k. \textbf{Left:} Proportions of trajectories corresponding to the three motion patterns in the train/validation/test splits. \textbf{Middle:} Distributions of trajectory lengths and geodesic distances to the sound source across all frames in the dataset. \textbf{Right:} Representative trajectories for each motion pattern, with audios shown as binaural spectrograms.}
   \label{fig:dataset}
\end{figure*}

In order to simulate the aforementioned environment, we describe an AVWM as follows:
\begin{equation}
\hat{o}_{t+\Delta t}\sim p_{\theta}\!\left(o_{t+\Delta t} \mid o_{t-m+1:t}, a_{t \rightarrow t+\Delta t}, \Delta t \right),
\end{equation}
where $o_{t-m+1:t}$ denotes the current observation along with its preceding frames, collectively referred to as the $m$-frame context.
$a_{t \rightarrow t+\Delta t}$ denotes the relative motion condition from the current frame $o_t$ to the target frame $o_{t+\Delta t}$, parameterized in our implementation by the displacement between the start and target positions together with the net heading change over the interval $[t, t+\Delta t]$.
$\Delta t \in [T_{\text{min}}, T_{\text{max}}]$ specifies the temporal offset measured in frames.
Since the target frame at time $t+\Delta t$ is determined by the overall motion between the current and target frames together with the time interval $\Delta t$, we use $a_{t \rightarrow t+\Delta t}$ for skip-step prediction.

The idea of conducting skip-step prediction by varying the temporal offset $\Delta t$ follows~\cite{bar2025navigation}, allowing the model not only to perform next-frame prediction.
Such temporal abstraction enables the world model to learn how the environment evolves over different time horizons, thereby improving its understanding of long-term spatio-temporal dependencies and underlying physical dynamics.

\subsection{AVW-4k Dataset}
\label{subsec:dataset}

To train and evaluate our model, we construct AVW-4k, a controlled audio-visual interaction benchmark collected in simulated indoor environments.

\noindent\textbf{Data Collection.}
Data collection is conducted within the simulation environment built upon the Matterport3D dataset~\cite{chang2017matterport3d} and the SoundSpaces 2.0 simulator~\cite{chen2022soundspaces}.
This allows for physically accurate sound propagation with frequency-dependent acoustic effects such as reflection, absorption, and reverberation.

During data collection, an agent continuously moves within the simulated 3D scenes while recording synchronized auditory and visual observations at each timestep.
Each environment contains a sound source at a known location, playing a looping ringtone.

This controlled setup is a deliberate benchmark design for isolating action-conditioned audio-visual dynamics.
  By anchoring the sound source in the scene and using a consistent source signal, the benchmark minimizes confounding
  variation from the emission process, making changes in the observations primarily attributable to agent motion,
  viewpoint, scene geometry, and sound propagation.
  As a result, the task directly evaluates whether a model can learn physically consistent coupling among egomotion,
  binaural acoustics, and synchronized visual change, rather than relying on incidental correlations from unconstrained
  audio-visual recordings.
  Such control is essential for attributing prediction errors to failures in world-dynamics modeling rather than to
  uncontrolled changes in the sound source.

The agent's trajectories are randomly sampled from three motion patterns: \textit{Source-approaching}, \textit{Source-receding}, and \textit{Source-independent}.
All trajectories are ensured to be navigable and maintain a reasonable distance from the source to prevent silent recordings.

Each frame consists of an egocentric RGB image with a resolution of $128 \times 128$ pixels and a 0.15-second binaural audio segment sampled at 16 kHz, strictly time-aligned across modalities.
The action space of AVW-4k includes four actions: move forward by 0.15 m, turn left by $10^\circ$, turn right by $10^\circ$, and stop.
Each sample contains no more than 500 frames.

\noindent\textbf{Dataset Composition.}
The final AVW-4k dataset contains approximately 30 hours of audio-visual data, spanning 76 indoor scenes and 4,500 trajectories. 
Trajectory lengths vary depending on navigation paths and acoustic configurations.
The dataset is divided into training, validation, and test splits with a ratio of 6 : 1 : 2, where the environments across subsets are non-overlapping, ensuring a fair evaluation of the model's cross-scene generalization ability. See Figure~\ref{fig:dataset} for dataset statistics and examples.

AVW-4k uniquely combines synchronized audio-visual streams, physically consistent acoustics, and low-level action labels for controlled embodied prediction; see the supplementary materials for comparison with existing benchmarks.

\section{Methodology}
\label{sec:method}

To implement the aforementioned AVWM, we design an integrated framework with two key innovations: the Audio-Visual Conditional Diffusion Transformer (AV-CDiT) featuring a novel modality expert architecture, and a stagewise training strategy.

\begin{figure*}[t]
  \centering
   \includegraphics[width=0.9\textwidth]{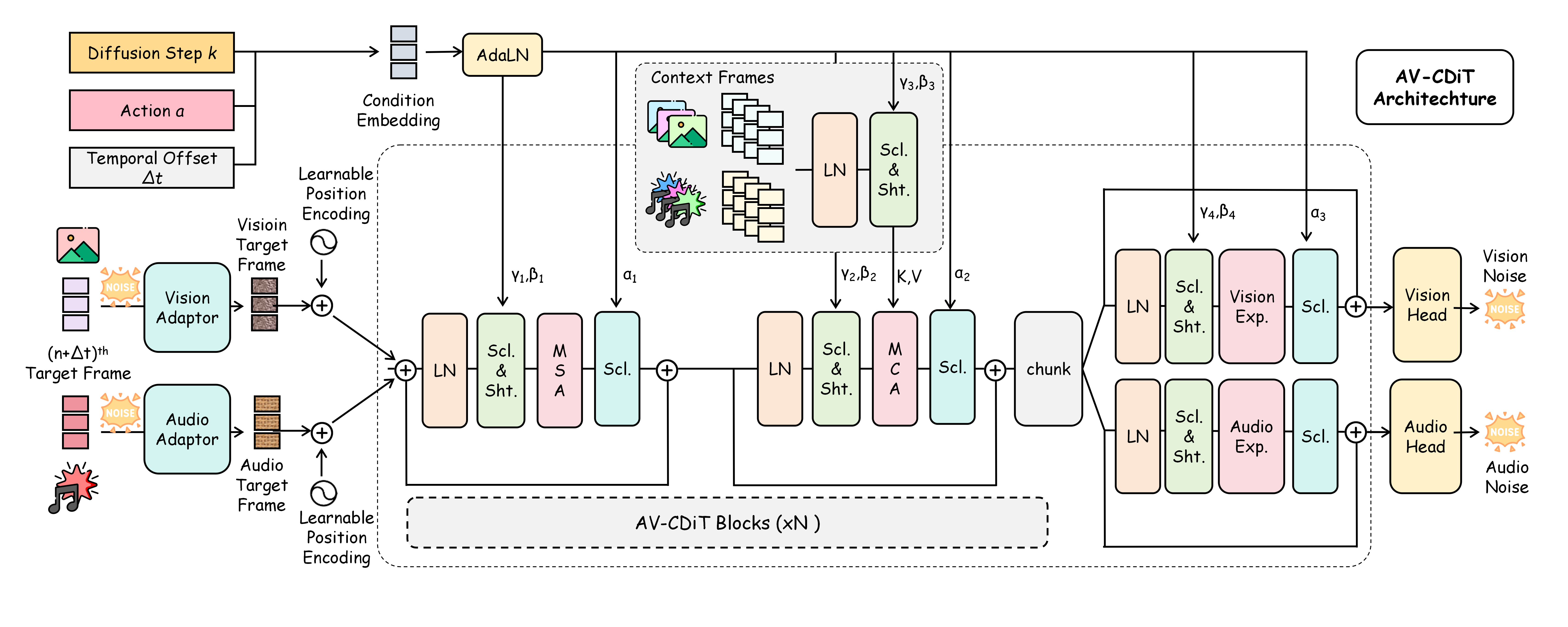}
   \caption{Overview of the proposed AV-CDiT architecture.}
   \label{fig:architecture}
\end{figure*}

\subsection{Audio-Visual Conditional Diffusion Transformer}
\label{subsec:architecture}

\noindent\textbf{Model Architecture.}
AV-CDiT is a temporally autoregressive Transformer model capable of generating both visual and auditory frames.
Building upon prior advances such as CDiT~\cite{bar2025navigation}, which was originally developed for unimodal visual world models, AV-CDiT extends the diffusion framework to support action-conditioned multimodal generation through a novel architecture, illustrated in Figure~\ref{fig:architecture}.

AV-CDiT first employs two pre-trained and frozen encoders to encode visual frames $o^v$ and auditory segments $o^a$, obtaining their respective latent representations $z^v$ and $z^a$, similar in spirit to latent diffusion models~\cite{rombach2022high}.
Subsequently, latent representations from both modalities are projected into a shared latent space through separate adaptor layers for alignment.
Finally, the visual and auditory tokens are concatenated to form the target sequence, which is noised during diffusion.
\begin{equation}
X_{t+\Delta t} = [h_{t+\Delta t}^v : h_{t+\Delta t}^a].
\end{equation}

This target sequence is subsequently fed into a stack of AV-CDiT blocks for multimodal diffusion-based generation.
Similar to the target sequence, the context sequence is constructed by concatenating context frames from both visual and auditory modalities.
\begin{equation}
C_{t-m+1:t} = [[h_{t-m+1}^v: h_{t-m+1}^a], \ldots, [h_t^v: h_t^a]].
\end{equation}

The conditioning embedding is generated following the CDiT formulation.
The relative motion condition $a_{t \rightarrow t+\Delta t}$, temporal offset $\Delta t$, and diffusion timestep $k$ are encoded via sinusoidal features and two-layer MLPs to obtain $\psi_a$, $\psi_{\Delta t}$, and $\psi_k$, which are summed into a single conditioning vector $c_t$.
This vector is then injected through an AdaLN~\cite{xu2019understanding} module to produce scale and shift parameters that modulate Layer Normalization and attention outputs, enabling conditional control during diffusion.

In each AV-CDiT Block, both visual and auditory tokens share common multi-head self-attention and cross-attention layers to enable joint modeling within a unified attention space.
The self-attention layer facilitates cross-modal temporal coherence and semantic alignment, while the cross-attention layer conditions the fused target tokens on the context representations $C_{t-m+1:t}$, allowing each modality to exploit complementary information during generation.

\noindent\textbf{Modality Experts.}
Inspired by previous expert network architectures~\cite{bao2022vlmo, wu2025janus}, each AV-CDiT block incorporates modality experts in the feed-forward layer that assign independent nonlinear mappings to each modality.
After shared self-attention and cross-attention operations, the target sequence is divided into visual and auditory groups processed by corresponding modality expert sub-networks, and their outputs are concatenated to reconstruct the unified sequence.
This design prevents visual dominance from hindering auditory representation learning when leveraging visually pretrained models, enabling the auditory pathway to learn in a coordinated manner alongside the visual modality, which works in conjunction with the stagewise training strategy introduced later.
After multiple stacked blocks, visual and auditory tokens are decoded through their respective output heads to predict noise, and the denoised tokens are then decoded into future observations.

\noindent\textbf{Diffusion Training.}
We follow the DDPM~\cite{ho2020denoising} formulation to model future multimodal generation as a synchronized diffusion process across both visual and auditory modalities.
Given the clean latent targets $X_{t+\Delta t}^v$ and $X_{t+\Delta t}^{a}$, Gaussian noise is independently injected into each modality according to a shared noise schedule $\{\bar{\alpha}_k\}_{k=1}^K$, while preserving their intrinsic correlation in the clean space:
\begin{equation}
\begin{aligned}
X_{t+\Delta t}^{v,(k)} &= \sqrt{\bar{\alpha}_k}\,X_{t+\Delta t}^v + \sqrt{1-\bar{\alpha}_k}\,\epsilon_v,\\
X_{t+\Delta t}^{a,(k)} &= \sqrt{\bar{\alpha}_k}\,X_{t+\Delta t}^{a} + \sqrt{1-\bar{\alpha}_k}\,\epsilon_{a},
\end{aligned}
\end{equation}
where $\epsilon_v,\,\epsilon_{a} \!\sim\! \mathcal{N}(0,I)$ are modality-specific Gaussian noises.
This formulation assumes conditionally independent forward noising processes for the two modalities.
Meanwhile, the multimodal dependency is retained in the reverse process through a unified denoising network.

\begin{figure*}[t]
  \centering
   \includegraphics[width=0.92\textwidth]{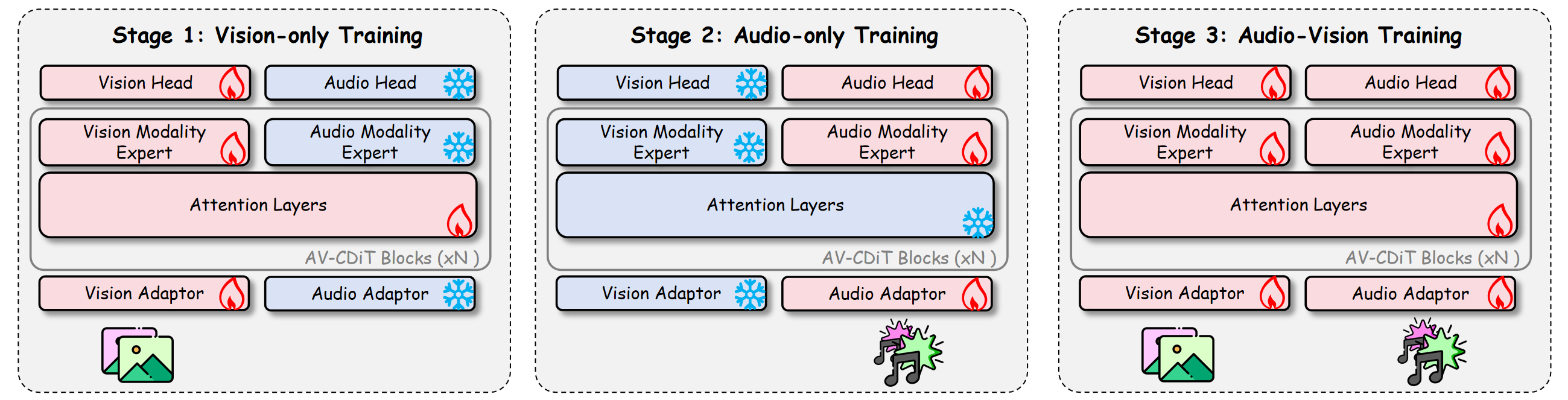}
   \caption{Illustration of the stagewise training strategy used for AV-CDiT.}
   \label{fig:training}
\end{figure*}

The reverse process is parameterized by AV-CDiT, which jointly predicts the injected noises $(\hat{\epsilon}_v, \hat{\epsilon}_{a})$ in a single forward pass.

\noindent\textbf{Training Objective.}
AV-CDiT adopts the standard $\epsilon$-prediction objective, estimating the noise added at each step rather than directly reconstructing clean targets.
The loss is defined as:
\begin{equation}
\mathcal{L}_{\text{simple}} =
\mathbb{E}_{k,\epsilon_v,\epsilon_{a}}
\!\left[
\|\hat{\epsilon}_v-\epsilon_v\|_2^2 +
\|\hat{\epsilon}_{a}-\epsilon_{a}\|_2^2
\right].
\end{equation}
Following \cite{peebles2023scalable}, we predict the covariance matrix of the noise and supervise it with the variational lower-bound loss $\mathcal{L}_{\text{vb}}$~\cite{nichol2021improved}, yielding the final objective
$\mathcal{L} = \mathcal{L}_{\text{simple}} + \lambda_{\text{vb}}\mathcal{L}_{\text{vb}}$.

\subsection{Stagewise Training Strategy}
\label{subsec:stages}

Inspired by previous work on image-text multimodal joint pre-trainings~\cite{bao2022vlmo, wu2025janus}, we adopt a three-stage training strategy to ensure stable optimization and efficient convergence of the audio-visual multimodal diffusion model, as illustrated in Figure~\ref{fig:training}.

In the first stage, the model is trained on vision-only data to learn spatial-temporal representations by jointly optimizing the self-attention layers, cross-attention layers, visual modality experts, visual adaptor layer, and visual head.
In the second stage, only the auditory modality experts, auditory adaptor layer, and auditory head are fine-tuned on audio-only data while freezing all other components, including shared attention modules and vision-related layers, to preserve previously acquired visual capabilities and prevent strong visual priors from suppressing auditory representation learning.
In the third stage, visual and auditory inputs are concatenated and the entire network undergoes end-to-end fine-tuning on synchronized audio-visual data, allowing all parameters to be jointly optimized for deeper multimodal fusion and improved consistency across generated visual frames and audio segments within the unified temporal diffusion process.

\section{Experiments and Results}
\label{sec:experiments}

\subsection{Experimental Setting}

\noindent\textbf{Evaluation Metrics.}
We evaluate AVWM along two axes: modality-level fidelity and physically grounded multisensory consistency in our controlled setting.
  For fidelity, we use LPIPS~\cite{zhang2018unreasonable}, DreamSim~\cite{fu2023dreamsim}, and PSNR for vision, and Log-Spectral Distance
  (LSD)~\cite{erell1990estimation} and spectral SSIM~\cite{wang2004image} for audio.
  For consistency, we report three motion-sensitive proxy metrics.
  ILD Error measures the absolute deviation between predicted and ground-truth interaural level differences, reflecting binaural spatial consistency.
  AV-Lag Error and AV-Corr Error are computed from cross-correlation between a visual motion-energy sequence and an audio-energy sequence, and measure the errors in peak lag and peak correlation, respectively, thereby assessing cross-modal temporal coupling.
  For AV-Lag Error and AV-Corr Error, we use the same ground-truth visual reference for generated and ground-truth audio to avoid the degenerate case where generated audio and generated video appear consistent simply because both deviate from the true dynamics in a correlated way; visual fidelity is evaluated separately.
  Full formulas are provided in the supplementary materials.

\noindent\textbf{Training Configuration.}
We initialize AV-CDiT from a pretrained NWM model~\cite{bar2025navigation} fine-tuned on AVW-4k, using a Stable Diffusion~\cite{blattmann2023stable} VAE for visual encoding and a SoundStream~\cite{zeghidour2021soundstream} tokenizer for audio encoding; all experiments are conducted on eight NVIDIA A100 GPUs.
Full encoder details and training hyperparameters are provided in the supplementary materials.

\noindent\textbf{Inference Mode.}
For each experimental configuration, we use \textbf{fixed-step} generation mode for the main evaluation.
Given a temporal offset $\Delta t$, the model directly predicts the target frame at $t+\Delta t$; we sweep $\Delta t$ over a 16-frame horizon.
In addition, we conduct supplementary experiments with \textbf{rollout} generation, where predictions are made autoregressively by feeding each generated frame back as input until reaching $t+16$. Results under the rollout setting are reported in the supplementary materials. All experimental results are evaluated for statistical significance using t-tests ($p < 0.05$).

\subsection{Comparison with Baselines}

  \noindent\textbf{Setup.}
  Since no existing world models are currently capable of synchronized audio-visual future prediction, we compare against representative
  baselines under two complementary settings.
  First, we consider \emph{modality-factorized} baselines, motivated by the fact that existing world models are typically unimodal and do not
  jointly model future audio and video.
  We therefore combine strong visual world models with a separate audio predictor to form audio-visual counterparts.
  Specifically, \textbf{DIAMOND}~\cite{alonso2024diffusion} and \textbf{NWM}~\cite{bar2025navigation} serve as the visual world models,
  while \textbf{AudioLDM}~\cite{liu2023audioldm} is used as the audio generator.
  To adapt AudioLDM to our setting, we encode the action vectors with a lightweight MLP, inject them into its conditioning pathway, and then
  fine-tune the model on AVW-4k.
  These models are combined as DIAMOND+AudioLDM and NWM+AudioLDM.

  \begin{table*}[t]
  \centering
  \small
  \setlength{\tabcolsep}{4pt}
  \begin{tabular}{@{}lcccccccc@{}}
  \toprule
  \textbf{Model} &
  \multicolumn{3}{c}{\textbf{Vision Fidelity}} &
  \multicolumn{2}{c}{\textbf{Audio Fidelity}} &
  \multicolumn{3}{c}{\textbf{Physical Consistency}} \\
  \cmidrule(lr){2-4}
  \cmidrule(lr){5-6}
  \cmidrule(l){7-9}
  &
  \textbf{LPIPS}$\downarrow$ &
  \textbf{DreamSim}$\downarrow$ &
  \textbf{PSNR}$\uparrow$ &
  \textbf{LSD}$\downarrow$ &
  \textbf{SSIM}$\uparrow$ &
  \textbf{ILD}$\downarrow$ &
  \textbf{AV-Lag}$\downarrow$ &
  \textbf{AV-Corr}$\downarrow$ \\
  \midrule
  DIAMOND+AudioLDM & $0.626$ & $0.447$ & $13.849$ &
  \multirow{2}{*}{$2.593$} &
  \multirow{2}{*}{$0.485$} &
  \multirow{2}{*}{$22.894$} &
  \multirow{2}{*}{$1.678$} &
  \multirow{2}{*}{$0.279$} \\
  NWM+AudioLDM     & $0.382$ & $0.258$ & $16.539$ &
  & & & & \\
  AVDiT+action     & $0.526$ & $0.443$ & $14.443$ &
  $1.252$ & $0.328$ & $3.003$ & $1.665$ & $0.294$ \\
  \textbf{AV-CDiT (Ours)} & $0.382$ & $0.255$ & $16.504$ &
  $1.311$ & $0.547$ & $2.471$ & $1.218$ & $0.194$ \\
  \bottomrule
  \end{tabular}
  \caption{Comparison with modality-factorized and joint audio-visual generation baselines. For DIAMOND+AudioLDM and NWM+AudioLDM, the audio branch is the same fine-tuned AudioLDM model, so the audio-fidelity and audio-visual consistency metrics are shared; the difference between the two rows comes from the visual branch.}
  \label{tab:baselines}
  \end{table*}

  Second, we consider a \emph{joint audio-visual generation} baseline.
  Here we adopt \textbf{AVDiT}~\cite{kim2024versatile}, which jointly models video and audio generation in a unified architecture rather than
  decomposing them into separate unimodal predictors.
  However, the original AVDiT is not designed as an action-conditioned world model.
  To adapt it to our setting, we inject the agent action as an additional global conditioning signal via a lightweight MLP, and fuse it into
  the model's conditioning pathway during denoising.
  This preserves the original AVDiT backbone while enabling low-level action-conditioned future prediction.

  Recent open-domain audio-visual generation models are also relevant related work, but they are not directly comparable under our setting,
  since many are not designed for low-level action control in embodied rollouts and primarily optimize semantic or perceptual alignment rather
  than physically grounded consistency under agent actions.
  We therefore do not include them as direct baselines in the present comparison.
  Detailed baseline implementations, including conditioning interfaces, preprocessing, and fine-tuning settings, are provided in the
  supplementary material.

\noindent\textbf{Results.}
  As shown in Table~\ref{tab:baselines}, AV-CDiT achieves the best overall performance across both fidelity and physical-consistency metrics.
  On the visual side, AV-CDiT matches the strong visual-only NWM baseline on LPIPS and attains slightly better DreamSim, while also modeling synchronized binaural audio in the same action-conditioned process.
  For audio, AV-CDiT substantially improves over the factorized AudioLDM-based baselines and achieves the best spectral SSIM, while AVDiT+action obtains slightly lower LSD but much worse visual and consistency scores.
  Most importantly, AV-CDiT clearly outperforms all baselines on ILD Error, AV-Lag Error, and AV-Corr Error, indicating better preservation of binaural spatial cues and tighter audio-visual temporal coupling.
  These results suggest that neither modality stitching nor naive joint audio-visual generation is sufficient to capture physically grounded cross-modal dynamics under action control.

\subsection{Ablations}
\label{subsec:ablation}

\noindent\textbf{Setup.}
We conduct ablation studies to verify the effects of the modality experts and the stagewise training strategy.
Specifically, we report the generative performance under two ablated settings.
In the first setting, the modality experts are removed and replaced with a shared feed-forward network, where the audio and visual token sequences are jointly processed.
In the second setting, the modality experts are retained, but the model is directly fine-tuned on both modalities simultaneously.

\begin{table*}[t]
  \centering
  \small
  \setlength{\tabcolsep}{4pt}
  \begin{tabular}{@{}lcccccccc@{}}
  \toprule
  \textbf{Model} &
  \multicolumn{3}{c}{\textbf{Vision Fidelity}} &
  \multicolumn{2}{c}{\textbf{Audio Fidelity}} &
  \multicolumn{3}{c}{\textbf{Physical Consistency}} \\
  \cmidrule(lr){2-4}
  \cmidrule(lr){5-6}
  \cmidrule(l){7-9}
  &
  \textbf{LPIPS}$\downarrow$ &
  \textbf{DreamSim}$\downarrow$ &
  \textbf{PSNR}$\uparrow$ &
  \textbf{LSD}$\downarrow$ &
  \textbf{SSIM}$\uparrow$ &
  \textbf{ILD}$\downarrow$ &
  \textbf{AV-Lag}$\downarrow$ &
  \textbf{AV-Corr}$\downarrow$ \\
  \midrule
  w/o modality experts &
  $0.378$ & $0.253$ & $16.536$ &
  $1.366$ & $0.485$ &
  $2.737$ & $1.618$ & $0.236$ \\
  w/o stagewise training &
  $0.380$ & $0.253$ & $16.545$ &
  $1.371$ & $0.485$ &
  $2.771$ & $1.611$ & $0.238$ \\
  \textbf{AV-CDiT (Ours)} &
  $0.382$ & $0.255$ & $16.504$ &
  $1.311$ & $0.547$ &
  $2.471$ & $1.218$ & $0.194$ \\
  \bottomrule
  \end{tabular}
  \caption{Comparison with different ablated variants in fixed-step generation mode.}
  \label{tab:variants}
  \end{table*}

\begin{figure}[t]
  \centering
   \includegraphics[width=0.9\columnwidth]{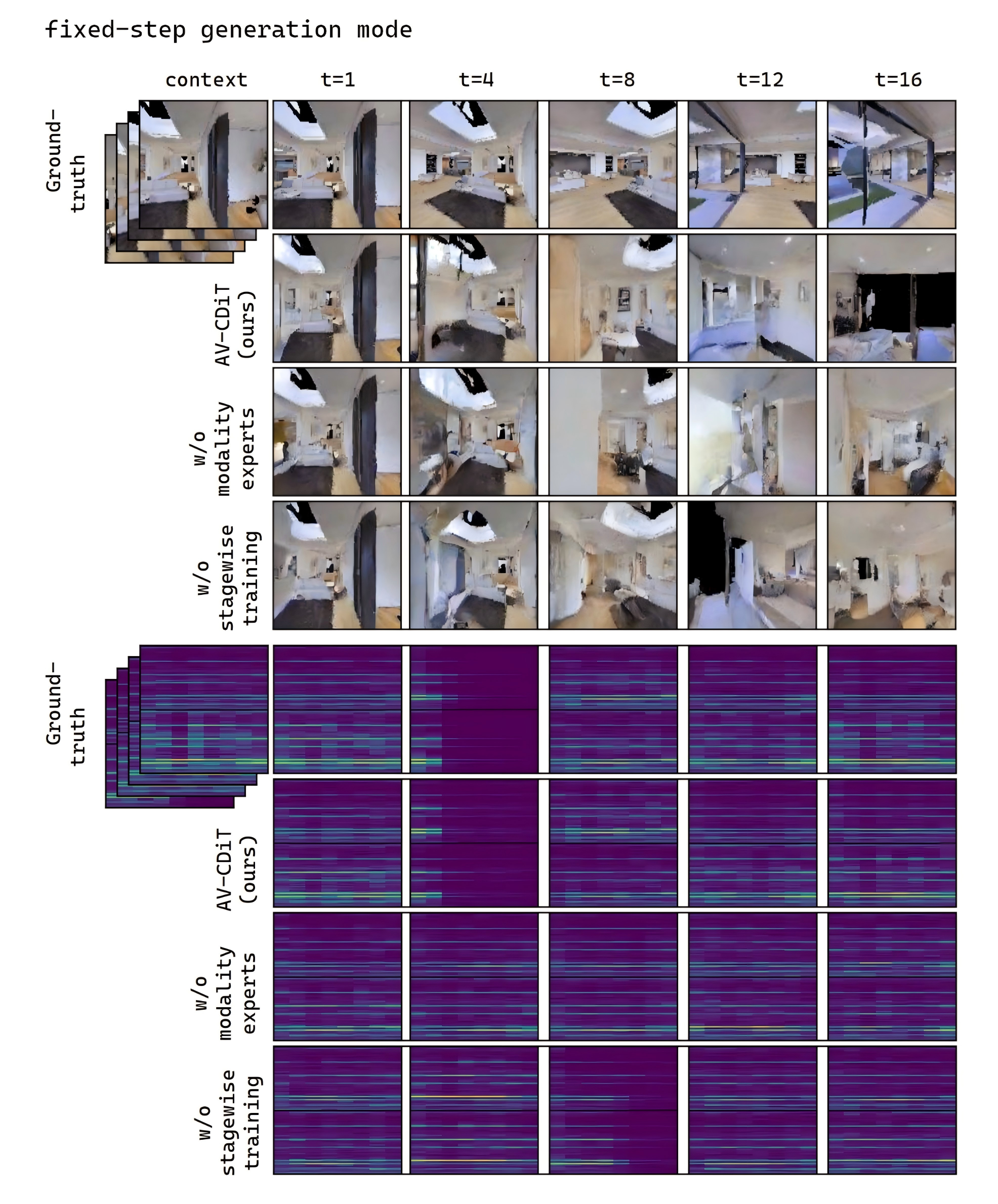}
   \caption{Qualitative analysis. This figure shows representative image and audio generation results of our model and two ablated variants under the fixed-step mode; additional examples are provided in supplementary materials.}
   \label{fig:ablations}
\end{figure}

  \noindent\textbf{Results.}
  As shown in Table~\ref{tab:variants}, all variants achieve similar visual fidelity, suggesting that the ablations do not substantially affect the visual prediction backbone.
  The main differences appear in audio fidelity and physical consistency.
  Compared with both ablated variants, the full AV-CDiT achieves better LSD and SSIM, and consistently reduces ILD Error, AV-Lag Error, and AV-Corr Error.
  These results indicate that modality experts and stagewise training are most beneficial for learning synchronized audio-visual dynamics, rather than merely improving per-modality appearance.
  Figure~\ref{fig:ablations} further provides qualitative comparisons under fixed-step generation, with additional examples in the supplementary materials.

\subsection{Planning with an Audio-Visual World Model}
  \noindent\textbf{Setup.}
  We evaluate the planning capability of AV-CDiT on continuous audio-visual navigation (continuous AV-Nav)~\cite{chen2022soundspaces}, where
  an agent navigates in SoundSpaces 2.0 environments built from Matterport3D scenes using egocentric RGB observations and binaural audio. An
  episode succeeds if the agent issues a \textit{stop} action within $1$m of the target sound source.

  Continuous AV-Nav uses the same action space as AVW-4k. We ensure that AV-CDiT training scenes are disjoint from the AV-Nav test scenes, so
  this experiment evaluates generalization of learned audio-visual dynamics to unseen indoor environments. For planning, we use a reward-aware
  AVWM lookahead planner built on a variant of AV-CDiT trained on a reward-annotated version of AVW-4k with an auxiliary progress-reward token.
  The reward is the reduction in geodesic distance to the target sound source and is used only for scoring candidate rollouts; reward
  annotation details and prediction checks are provided in the supplementary materials. Specifically, the visual and auditory latent tokens are concatenated with a broadcast
  reward token,
  enabling prediction of future observations together with the immediate progress reward.

  We use the resulting AVWM to improve a pretrained AV-Nav policy~\cite{chen2022soundspaces} through lookahead planning. At each timestep
  $t$, the agent samples $n$ candidate actions from the policy distribution. Each candidate is propagated through AVWM for a $k$-step rollout,
  producing predicted future observations and progress rewards. At each rollout step, the predicted progress reward is combined with the
  policy network's value estimate and discounted over time to compute a branch score. Beam search retains up to $B$ top-scoring action
  sequences, and the agent executes the first action of the highest-scoring sequence in the real environment.

  We report results for different beam widths $B$ and rollout lengths $k$, with $n=2$ candidate actions. Performance is measured by SPL,
  softSPL, number of actions (NA), and success normalized by action (SNA).

\noindent\textbf{Results.}
As shown in Table~\ref{tab:nav}, this reward-aware AVWM lookahead planner effectively enhances the agent's navigation performance under proper parameter configurations.
In particular, the improvement in navigation efficiency, reflected by a significant reduction in the number of actions (NA), is especially pronounced.
AVWM-based foresight enables the agent to evaluate multiple possible future outcomes before acting, leading to more informed and goal-directed decisions. The oracle results further indicate headroom for improving learned rollout quality.

  \begin{table}[t]
  \centering
  \footnotesize
  \setlength{\tabcolsep}{1mm}
  \begin{tabular}{@{}lllcccc@{}}
  \toprule
  \multicolumn{3}{@{}l}{Setting} & SPL$\uparrow$ & SoftSPL$\uparrow$ & NA$\downarrow$ & SNA$\uparrow$ \\
  \midrule
  \multicolumn{3}{@{}l}{AV-Nav Agent (baseline)} & $45.08$ & $55.09$ & $332.817$ & $6.44$ \\
  \midrule
  $B=3$ & $k=4$ & AVWM & $45.70$ & $56.54$ & $321.11$ & $6.67$ \\
        &       & \emph{Oracle WM} & \emph{51.02} & \emph{60.71} & \emph{300.35} & \emph{8.36} \\
        & $k=5$ & AVWM & $\mathbf{47.38}$ & $56.53$ & $\mathbf{317.19}$ & $\mathbf{7.09}$ \\
        &       & \emph{Oracle WM} & \emph{51.98} & \emph{61.44} & \emph{297.68} & \emph{8.53} \\
  \midrule
  $B=4$ & $k=4$ & AVWM & $46.46$ & $57.00$ & $317.99$ & $6.87$ \\
        &       & \emph{Oracle WM} & \emph{51.64} & \emph{60.58} & \emph{298.25} & \emph{8.48} \\
        & $k=5$ & AVWM & $46.74$ & $\mathbf{57.24}$ & $317.82$ & $6.84$ \\
        &       & \emph{Oracle WM} & \emph{50.50} & \emph{60.58} & \emph{301.52} & \emph{8.17} \\
  \bottomrule
  \end{tabular}
  \caption{Performance improvement of the continuous AV-Nav agent with AVWM. `Oracle WM' denotes an upper bound obtained by replacing the outcomes and reward from AVWM with the ground-truth environment feedback.}
  \label{tab:nav}
  \end{table}

\begin{figure}[t]
\centering
\includegraphics[width=0.8\columnwidth]{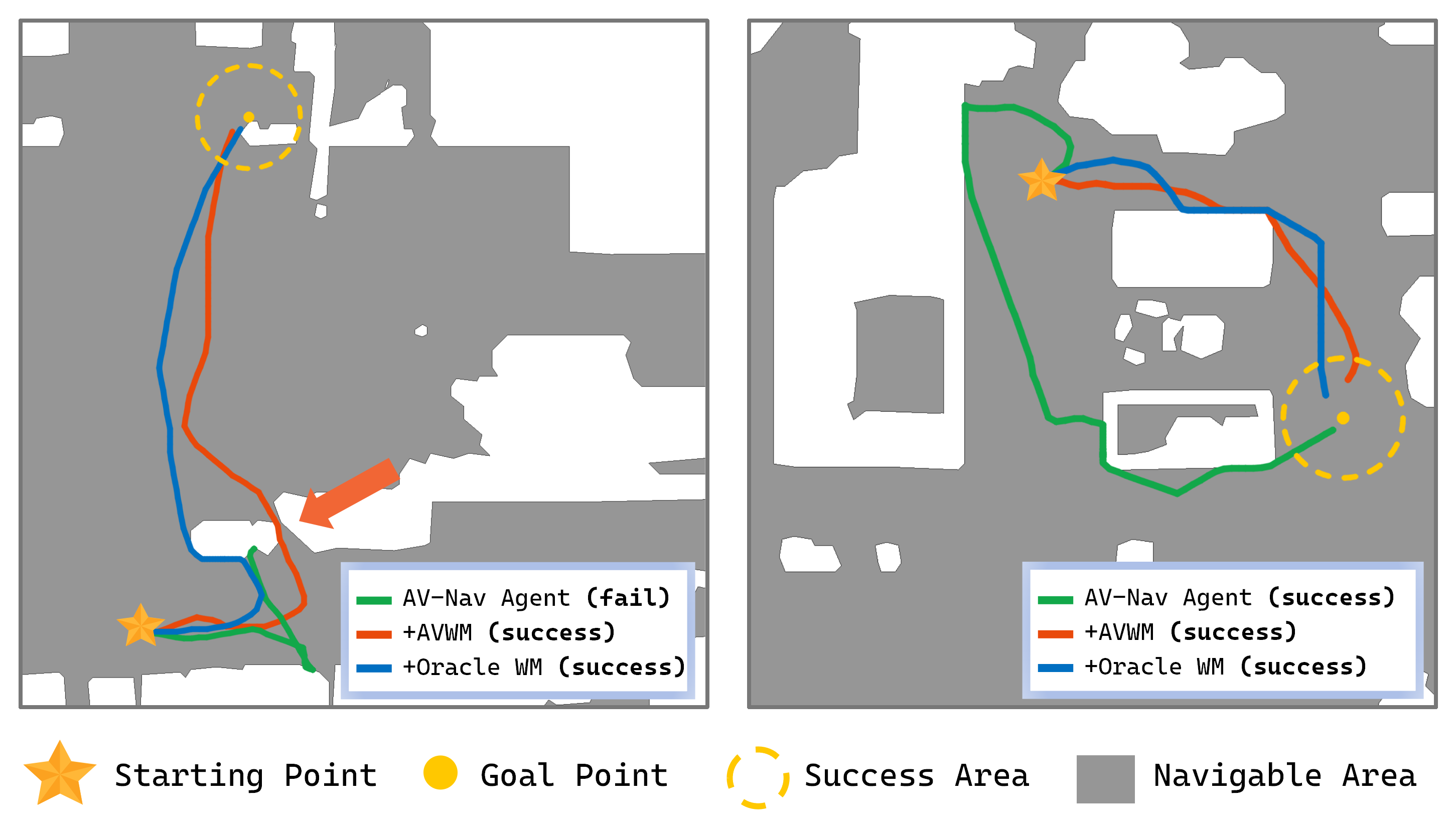} %
\caption{Comparison of navigation trajectories produced by three agents: the baseline AV-Nav agent (green), the AV-Nav agent enhanced with AVWM planning (red; $n=2$, $B=3$, $k=5$), and the upper-bound AV-Nav agent augmented with an oracle world model (blue).}
\label{fig:nav}
\end{figure}

  \section{Conclusion}

We introduced AVWM, a unified formulation for action-conditioned audio-visual world modeling, together with AVW-4k and AV-CDiT. Through modality experts and stagewise training, AV-CDiT achieves strong generation quality, physically grounded consistency, and improved downstream audio-visual navigation. The benchmark focuses on synthetic indoor environments with spatially anchored sound sources, motivated by the scarcity of real-world datasets with synchronized binaural audio-visual streams and precise action labels. Supplementary experiments further test more diverse sound-source categories. Future
  work will extend AVWM to richer sound events, moving sources, broader acoustic variability, and additional transformer-based generative
  backbones.

\newpage
\clearpage

\twocolumn[
\begin{center}
{\LARGE \bfseries Supplementary Materials\par}
\vspace{1em}
\end{center}
]

\section{Comparison with Analogous Datasets}
\label{app:dataset_comparison}

To contextualize AVW-4k among analogous datasets and benchmarks, we compare it with several analogous datasets in Table~\ref{tab:dataset_comparison}.
The evaluation dimensions used in our comparison are explained as follows.
\begin{itemize}
    \item \textbf{Vision (Vis.) \& Audio (Aud.)} indicates whether the dataset provides synchronized visual and auditory streams for paired multisensory modeling.
    \item \textbf{Real/Syn.} distinguishes between real-world captured data (Real) and synthetic data generated in simulation (Syn.).
    \item \textbf{Physical Consistency} indicates whether the audio stream is physically grounded in the visible scene and its dynamics, rather than merely accompanying the video; datasets are considered inconsistent when the audio contains substantial components not causally tied to the observed environment, such as voiceovers, commentary, background music, or mixed external speech.
    \item \textbf{Action Type} characterizes the granularity of control signals available, categorized as \textbf{None}, \textbf{Text Prompt} (e.g., high-level descriptions such as ``washing dishes''), and \textbf{Low-Level Action} (e.g., low-level embodied actions such as displacement and rotation).
    \item \textbf{Suitability for AVWM} indicates whether the dataset provides the synchronized multisensory observations, physical audio-visual grounding, and action supervision required for action-contingent audio-visual future prediction.
\end{itemize}

  \begin{table*}[t]
  \centering
  \scriptsize
  \setlength{\tabcolsep}{2pt}
  \renewcommand{\arraystretch}{0.95}
  \begin{tabular}{@{}lcccccc@{}}
  \toprule
  \textbf{Dataset} & \textbf{Vis.} & \textbf{Aud.} & \shortstack[c]{\textbf{Physical}\\\textbf{Cons.}} & \shortstack[c]
  {\textbf{Real/}\\\textbf{Syn.}} &
  \shortstack[c]{\textbf{Action}\\\textbf{Type}} & \shortstack[c]{\textbf{AVWM}\\\textbf{Suit.}} \\
  \midrule
  AudioSet~\citep{gemmeke2017audio} & \checkmark & \checkmark & \ding{55} & Real & None & \ding{55} \\
  Landscape~\citep{lee2022sound} & \checkmark & \checkmark & \checkmark & Real & None & \ding{55} \\
  EPIC-KITCHENS~\citep{damen2020epic} & \checkmark & \checkmark & \checkmark & Real & Text Prompt & \ding{55} \\
  Ego4D-Sounds~\citep{chen2024action2sound} & \checkmark & \checkmark & \checkmark & Real & Text Prompt & \ding{55} \\
  SCAND~\citep{karnan2022socially} & \checkmark & \ding{55} & N/A & Real & Low-Level Action & \ding{55} \\
  RoboNet~\citep{dasari2019robonet} & \checkmark & \ding{55} & N/A & Real & Low-Level Action & \ding{55} \\
  Meta-World~\citep{yu2020meta} & \checkmark & \ding{55} & N/A & Syn. & Low-Level Action & \ding{55} \\
  CSGO~\citep{pearce2022counter} & \checkmark & \ding{55} & N/A & Syn. & Low-Level Action & \ding{55} \\
  MineRL~\citep{guss2019minerl} & \checkmark & \ding{55} & N/A & Syn. & Low-Level Action & \ding{55} \\
  PLAICraft~\citep{he2025plaicraft} & \checkmark & \checkmark & \ding{55} & Syn. & Low-Level Action & \ding{55} \\
  \textbf{AVW-4k (Ours)} & \checkmark & \checkmark & \checkmark & Syn. & Low-Level Action & \checkmark \\
  \bottomrule
  \end{tabular}
  \caption{Comparison of AVW-4k with analogous datasets and benchmarks. Vis., Aud., Physical Cons., Real/Syn., and AVWM Suit. denote vision, audio, physical consistency, real or synthetic data, and suitability for audio-visual world modeling, respectively.}
  \label{tab:dataset_comparison}
\end{table*}

\noindent\textbf{Dataset Descriptions.}
\begin{itemize}
    \item \textbf{AudioSet} is a large-scale audio event dataset curated by Google researchers through manual annotation of 10-second YouTube segments.
    Primarily designed for passive recognition, it is unsuitable for AVWM because its audio often contains post-production dubbing, background music, or other components that are not physically grounded in the visible scene, and it lacks low-level action labels required for action-contingent modeling.
    \item \textbf{Landscape} dataset is a high-resolution collection of natural scenes, totaling approximately 26 hours; while it offers superior physical AV-consistency by accurately capturing environmental acoustics, it is a passive observation dataset that lacks any action labels.
    \item \textbf{EPIC-KITCHENS} is a large-scale egocentric video dataset featuring 55 hours of unscripted daily activities in kitchens, providing textual annotations of action segments (e.g., \textit{`Stir chicken.'}) and object bounding boxes based on participants' post-recording narrations, but without precise low-level actions.
    \item \textbf{Ego4D-Sounds} consists of 1.2 million curated clips from Ego4D~\citep{grauman2022ego4d} and EPIC-KITCHENS, specifically designed to strengthen the correspondence between foreground actions and audio by disentangling ambient background sounds; however, the action data remains limited to post-annotated textual descriptions of activities, lacking the precise, low-level motion control.
    \item \textbf{SCAND} is a large-scale, first-person-view dataset for socially compliant navigation, featuring 25 miles of human tele-operated driving demonstrations; while it provides precise low-level action data, including joystick commands and odometry, the dataset lacks synchronized audio streams.
    \item \textbf{RoboNet} is a large-scale multi-robot interaction dataset featuring 162,000 trajectories and 15 million video frames, focusing on predicting future video frames conditioned on precise action sequences; however, the total absence of the auditory modality makes it unsuitable for AVWMs.
    \item \textbf{Meta-World} is an open-source simulated benchmark for robotic manipulation; the environment provides only visual observations and robotic actions, with a total absence of audio signals.
    \item \textbf{CSGO} dataset is derived from Counter-Strike: Global Offensive, featuring 95 hours of human gameplay demonstrations with precise action labels aligned with visual game frames, but does not contain any synchronized audio streams.
    \item \textbf{MineRL} is a large-scale synthetic dataset based on Minecraft, featuring over 60 million state-action pairs generated from human demonstrations; its focus is primarily on visual pixels and game states without synchronized audio streams.
    \item \textbf{PLAICraft} is a large-scale, multiplayer Minecraft interaction dataset featuring video, audio, and precise action sequences; although it records low-level control data, its audio stream is contaminated by inseparable, mixed-in player microphone speech, which introduces significant components that are physically ungrounded in the visual stream.
\end{itemize}

\section{Trajectory Motion Patterns}
\label{app:motion_patterns}

AVW-4k trajectories are sampled from three motion patterns designed to cover different relationships between the agent path and the stationary sound source.
\begin{itemize}
    \item \textbf{Source-approaching} trajectories begin from a random navigable position and move toward the source.
    \item \textbf{Source-receding} trajectories begin near the source and move toward a random goal position.
    \item \textbf{Source-independent} trajectories move between two random navigable positions independently of the source location.
\end{itemize}
These patterns provide a mixture of paths in which the source becomes progressively more salient, less salient, or changes only incidentally with egomotion.

\section{Details of Evaluation Metrics}
\label{app:metrics}

We evaluate AVWM from two complementary perspectives: modality-level fidelity and operational proxies for physically grounded multisensory consistency.
Modality-level fidelity metrics provide indirect evidence that the generated observations preserve the structure of real sensory signals, while the consistency metrics are designed to measure whether the generated audio remains spatially and temporally coupled to the underlying visual dynamics in our controlled benchmark.

\noindent\textbf{Visual Fidelity.}
For visual prediction, we follow prior unimodal visual world models~\citep{bar2025navigation, wu2025rlvr} and report
LPIPS~\citep{zhang2018unreasonable}, DreamSim~\citep{fu2023dreamsim}, and PSNR.
LPIPS and DreamSim evaluate perceptual similarity between predicted and ground-truth frames using deep visual features.
PSNR measures pixel-level reconstruction quality,
\begin{equation}
\mathrm{PSNR} = 10 \log_{10}\!\left(\frac{\mathrm{MAX}^2}{\mathrm{MSE}}\right),
\end{equation}
where $\mathrm{MAX}$ is the maximum pixel value and $\mathrm{MSE}$ is the mean squared error between the predicted and ground-truth images.

\noindent\textbf{Audio Fidelity.}
For audio prediction, we retain LSD~\citep{erell1990estimation} and spectral SSIM~\citep{wang2004image}.
In implementation, both generated and ground-truth audio clips are zero-padded to ensure compatibility with the VGGish feature extractor.
LSD measures frequency-domain deviation between the generated and ground-truth spectra:
\begin{equation}
\mathrm{LSD} = \frac{1}{T}\sum_{t=1}^{T}
\sqrt{\frac{1}{F}\sum_{f=1}^{F}
\left(\log S_t(f)-\log \hat{S}_t(f)\right)^2},
\end{equation}
where $S_t(f)$ and $\hat{S}_t(f)$ denote the ground-truth and predicted magnitude spectra at time index $t$ and frequency bin $f$.
Spectral SSIM measures structural similarity between predicted and reference spectrograms.
Since our model generates binaural audio, channel-wise fidelity metrics are computed independently on the left and right channels and then averaged.

\noindent\textbf{Interaural Spatial Consistency.}
To evaluate whether generated binaural audio preserves physically meaningful interaural cues in this setting, we compute \textbf{ILD Error} for each predicted audio segment.
The interaural level difference (ILD) is computed from channel-wise RMS energy in decibels:

\begin{equation}
  \begin{aligned}
  \mathrm{ILD}(x_L, x_R)
  &= 20\log_{10}\!\left(\mathrm{RMS}(x_L)\right) \\
  &\quad - 20\log_{10}\!\left(\mathrm{RMS}(x_R)\right).
  \end{aligned}
\end{equation}

with
\begin{equation}
\mathrm{RMS}(x) = \sqrt{\frac{1}{N}\sum_{n=1}^{N} x[n]^2}.
\end{equation}
The ILD error is defined as
\begin{equation}
\mathrm{ILD\ Error} = \left| \widehat{\mathrm{ILD}} - \mathrm{ILD}^{*} \right|.
\end{equation}

\noindent\textbf{Cross-Modal Temporal Consistency.}
To evaluate whether the generated audio evolves with the expected temporal coupling to visual dynamics under our benchmark design, we compute \textbf{AV-Lag Error} and \textbf{AV-Corr Error} from low-level visual motion and audio-energy sequences.

For the visual stream, we define a motion-energy sequence from consecutive grayscale frames:
\begin{equation}
v_t = \mathrm{mean}\!\left(\left| I_t - I_{t-1} \right|\right),
\end{equation}
where $I_t$ is the grayscale image at time step $t$.
For the audio stream, we first compute a short-time binaural envelope within each audio segment using the average RMS of the left and right channels over local analysis windows:
\begin{equation}
e_{t,k} = \frac{1}{2}\left(\mathrm{RMS}(x_{L,t,k}) + \mathrm{RMS}(x_{R,t,k})\right),
\end{equation}
where $k$ indexes the short-time analysis window inside segment $t$.
We then aggregate this envelope into a segment-level audio-energy sequence:
\begin{equation}
a_t = \frac{1}{K_t}\sum_{k=1}^{K_t} e_{t,k}.
\end{equation}

\begin{table*}[t]
\centering
\footnotesize
\setlength{\tabcolsep}{3pt}
\renewcommand{\arraystretch}{0.95}
\begin{tabular}{@{}lcccccccc@{}}
\toprule
\textbf{Model} &
\multicolumn{3}{c}{\textbf{Vision Fidelity}} &
\multicolumn{2}{c}{\textbf{Audio Fidelity}} &
\multicolumn{3}{c}{\textbf{Physical Consistency}} \\
\cmidrule(lr){2-4}
\cmidrule(lr){5-6}
\cmidrule(l){7-9}
&
\textbf{LPIPS}$\downarrow$ &
\textbf{DreamSim}$\downarrow$ &
\textbf{PSNR}$\uparrow$ &
\textbf{LSD}$\downarrow$ &
\textbf{SSIM}$\uparrow$ &
\textbf{ILD}$\downarrow$ &
\textbf{AV-Lag}$\downarrow$ &
\textbf{AV-Corr}$\downarrow$ \\
\midrule
Pretrained NWM & $0.628$ & $0.516$ & $11.828$ & $\emptyset$ & $\emptyset$ & $\emptyset$ & $\emptyset$ & $\emptyset$ \\
Training Stage 1 & $0.382$ & $0.258$ & $16.539$ & $\emptyset$ & $\emptyset$ & $\emptyset$ & $\emptyset$ & $\emptyset$ \\
Training Stage 1+2 & $\emptyset$ & $\emptyset$ & $\emptyset$ & $1.297$ & $0.531$ & $2.529$ & $1.249$ & $0.200$ \\
\textbf{Training Stage 1+2+3 (Ours)} & $0.382$ & $\mathbf{0.255}$ & $16.504$ & $1.311$ & $0.547$ & $2.471$ & $1.218$ & $0.194$ \\
\bottomrule
\end{tabular}
\caption{Performance under \textbf{fixed-step} generation mode during training stages. `$\emptyset$' marks that the model does not produce outputs for this modality at this stage.}
\label{tab:fixed-step-part1}
\end{table*}

\begin{table*}[t]
\centering
\footnotesize
\setlength{\tabcolsep}{3pt}
\renewcommand{\arraystretch}{0.95}
\begin{tabular}{@{}lcccccccc@{}}
\toprule
\textbf{Model} &
\multicolumn{3}{c}{\textbf{Vision Fidelity}} &
\multicolumn{2}{c}{\textbf{Audio Fidelity}} &
\multicolumn{3}{c}{\textbf{Physical Consistency}} \\
\cmidrule(lr){2-4}
\cmidrule(lr){5-6}
\cmidrule(l){7-9}
&
\textbf{LPIPS}$\downarrow$ &
\textbf{DreamSim}$\downarrow$ &
\textbf{PSNR}$\uparrow$ &
\textbf{LSD}$\downarrow$ &
\textbf{SSIM}$\uparrow$ &
\textbf{ILD}$\downarrow$ &
\textbf{AV-Lag}$\downarrow$ &
\textbf{AV-Corr}$\downarrow$ \\
\midrule
Pretrained NWM & $0.667$ & $0.557$ & $11.435$ & $\emptyset$ & $\emptyset$ & $\emptyset$ & $\emptyset$ & $\emptyset$ \\
Training Stage 1 & $0.414$ & $0.278$ & $15.941$ & $\emptyset$ & $\emptyset$ & $\emptyset$ & $\emptyset$ & $\emptyset$ \\
Training Stage 1+2 & $\emptyset$ & $\emptyset$ & $\emptyset$ & $1.628$ & $0.544$ & $2.679$ & $0.869$ & $0.324$ \\
\textbf{Training Stage 1+2+3 (Ours)} & $0.407$ & $0.272$ & $16.019$ & $1.620$ & $0.577$ & $2.654$ & $0.867$ & $0.322$ \\
\bottomrule
\end{tabular}
\caption{Performance under \textbf{rollout} generation mode during training stages. Settings follow Table~\ref{tab:fixed-step-part1}.}
\label{tab:rollout-part1}
\end{table*}

\begin{table*}[t]
\centering
\footnotesize
\setlength{\tabcolsep}{3pt}
\renewcommand{\arraystretch}{0.95}
\begin{tabular}{@{}lcccccccc@{}}
\toprule
\textbf{Model} &
\multicolumn{3}{c}{\textbf{Vision Fidelity}} &
\multicolumn{2}{c}{\textbf{Audio Fidelity}} &
\multicolumn{3}{c}{\textbf{Physical Consistency}} \\
\cmidrule(lr){2-4}
\cmidrule(lr){5-6}
\cmidrule(l){7-9}
&
\textbf{LPIPS}$\downarrow$ &
\textbf{DreamSim}$\downarrow$ &
\textbf{PSNR}$\uparrow$ &
\textbf{LSD}$\downarrow$ &
\textbf{SSIM}$\uparrow$ &
\textbf{ILD}$\downarrow$ &
\textbf{AV-Lag}$\downarrow$ &
\textbf{AV-Corr}$\downarrow$ \\
\midrule
w/o modality experts &
$0.403$ & $0.270$ & $16.013$ &
$1.668$ & $0.515$ &
$2.694$ & $0.885$ & $0.330$ \\
w/o stagewise training &
$0.404$ & $0.269$ & $16.089$ &
$1.614$ & $0.507$ &
$2.706$ & $0.892$ & $0.332$ \\
\textbf{AV-CDiT (Ours)} &
$0.407$ & $0.272$ & $16.019$ &
$1.620$ & $0.577$ &
$2.654$ & $0.867$ & $0.322$ \\
\bottomrule
\end{tabular}
\caption{Comparison with different ablated variants under \textbf{rollout} generation mode.}
\label{tab:rollout-part2}
\end{table*}

Using the ground-truth visual trajectory as reference, we compute the normalized cross-correlation between the motion sequence $v_t$ and the audio-energy sequence $a_t$ within a local lag window:
\begin{equation}
C(\tau) = \frac{1}{N_\tau}\sum_{t} \tilde{v}_t \,\tilde{a}_{t+\tau},
\qquad \tau \in [-L,L],
\end{equation}
where $\tilde{v}_t$ and $\tilde{a}_t$ denote standardized sequences and $L$ is the maximum lag.
The optimal lag and peak correlation are
\begin{equation}
\tau^{*} = \arg\max_{\tau \in [-L,L]} C(\tau),
\qquad
C^{*} = \max_{\tau \in [-L,L]} C(\tau).
\end{equation}

We compute these quantities for both generated and ground-truth audio using the same visual reference sequence.
The final errors are
\begin{equation}
\mathrm{AV\text{-}Lag\ Error} = \left| \hat{\tau}^{*} - \tau^{*}_{\mathrm{gt}} \right|,
\end{equation}
\begin{equation}
\mathrm{AV\text{-}Corr\ Error} = \left| \hat{C}^{*} - C^{*}_{\mathrm{gt}} \right|,
\end{equation}
where $(\hat{\tau}^{*}, \hat{C}^{*})$ are computed from the generated audio and $(\tau^{*}_{\mathrm{gt}}, C^{*}_{\mathrm{gt}})$ from the ground-truth audio.

Using the same ground-truth visual reference avoids the degenerate case where predicted audio and predicted video appear mutually consistent simply because both deviate from the true dynamics in a correlated way. Visual fidelity is evaluated separately.

\section{Training Configuration}
\label{app:training_config}

\noindent\textbf{Encoder Configuration.}
For the visual encoder, we use the Stable Diffusion~\citep{blattmann2023stable} VAE tokenizer, similar to that employed in DiT~\citep{peebles2023scalable} and CDiT.
For the auditory encoder, we trained a SoundStream~\citep{zeghidour2021soundstream} tokenizer on AVW-4k, similar to that employed in AVDiT~\citep{kim2024versatile}.
Notably, we remove one extraction block from the original SoundStream encoder-decoder architecture to better accommodate the shorter audio segments in our dataset.

\noindent\textbf{Training Details.}
For the implementation of the stagewise training strategy, we first fine-tune a pretrained NWM model~\citep{bar2025navigation} of base size on the visual subset of the AVW-4k dataset.
The model has been pre-trained on large-scale visual datasets and thus possesses certain visual reasoning capabilities.
The parameters of its CDiT blocks, including those of the self-attention, cross-attention, and feed-forward layers, are then used to initialize the corresponding layers of an AV-CDiT model of identical size, marking the completion of the first stage.
Subsequently, we continue fine-tuning through the remaining two stages.
The AV-CDiT model is trained with a context length of 4 frames.
During training, each sample randomly selects 4 different navigation goals within a temporal window of $\pm 16$ frames around the current timestep for prediction, following the training paradigm of CDiT.
We use the AdamW optimizer.
The learning rates are set to $1.6 \times 10^{-4}$, $8 \times 10^{-4}$, and $1.6 \times 10^{-4}$ for the three stages.
The total batch sizes for the three stages are $512$, $768$, and $128$, respectively.

\section{Baseline Implementations}
\label{app:baseline_impl}

\noindent\textbf{AVDiT + action.}
The original AVDiT is a joint audio-video latent diffusion Transformer trained with MoNL (mixture of noise levels), which enables a single model to learn diverse conditional generation tasks, including joint generation, cross-modal generation, and continuation.
Its basic design encodes audio and videbaselineo into latent spaces, maps them into a shared token sequence, models them jointly with a Transformer, and injects diffusion conditioning through timestep embeddings and AdaLN modulation.
In our experiments, AVDiT is used as a joint audio-visual generation baseline with minimal adaptation to the action-conditioned future prediction setting.
We preserve its joint Transformer backbone and MoNL-based diffusion formulation, and only adapt the input-output interface to context-conditioned future prediction.
Action and relative-time information are encoded as an additional global condition, which is added to the timestep-derived conditioning tokens before AdaLN modulation throughout the network.
This keeps the baseline close to the original AVDiT design while enabling low-level action-conditioned future observation prediction.

\noindent\textbf{DIAMOND.}
In our experiments, DIAMOND is used as the visual branch of the modality-factorized baselines.
We follow its original visual diffusion prediction formulation and train it on AVW-4k for future image prediction, where the inputs are past visual observations and action conditions, and the output is the future visual frame.
Other than adapting the data format and training resolution, we do not alter its core visual world-model design.

\noindent\textbf{NWM.}
NWM~\citep{bar2025navigation} is a controllable visual world model based on a Conditional Diffusion Transformer (CDiT), which predicts future visual observations from past observations and navigation actions.
In our experiments, NWM is used as the visual branch of the modality-factorized baselines.
We follow its original action-conditioned visual prediction formulation and fine-tune it on AVW-4k for future image prediction.
Other than adapting the data interface to our benchmark, we keep its core CDiT-based visual world-model design unchanged.

\noindent\textbf{AudioLDM.}
AudioLDM is a latent diffusion model for audio generation, operating in a compressed audio representation space and conditioning generation through feature modulation.
In this work, we use AudioLDM as the audio branch of the modality-factorized baselines, with a light adaptation of its conditioning interface for action-conditioned future audio prediction.
Specifically, we encode low-level action-related trajectory information into a conditioning vector and inject it through its original conditioning pathway, while keeping the latent diffusion backbone and audio generation formulation unchanged.
The model is initialized from pretrained weights and fine-tuned on AVW-4k audio data.

\section{Additional Experimental Results}
\label{app:rollout_results}

\subsection{Performance Evolution during Training Stages}
\label{app:training_stages}

\noindent\textbf{Setup\noindent\textbf{Results.} .} We report the generative performance across the complete training stages, including a pre-trained NWM without training on AVW-4k, AV-CDiT fine-tuned on images (after stage 1), on audio (after stage 2), and jointly on both modalities (after stage 3).

\noindent\textbf{Results.} As shown in Table~\ref{tab:fixed-step-part1} and Table~\ref{tab:rollout-part1}, the three-stage training strategy effectively enables the model to maintain strong generative performance across both visual and auditory modalities.
After the first stage, the model's visual generation quality improves substantially.
After the second stage, the model develops the ability to reason about the auditory modality and generate plausible auditory outputs.
After the third stage, the model not only retains the modality-specific knowledge learned in the previous stages, thus avoiding catastrophic forgetting, but also achieves better cross-modal alignment and joint representation learning, leading to further improvements in generation quality, particularly in the auditory modality.
Beyond these settings, we further investigated the effect of unfreezing all layers across all stages; this led to divergence in the second stage despite extensive hyperparameter tuning, which demonstrates that maintaining fixed attention in the second stage is vital to preserve pre-trained spatial knowledge and prevent catastrophic forgetting.

\subsection{Ablations under Rollout Generation Mode}

\noindent\textbf{Results.} Table~\ref{tab:rollout-part2} provides the rollout-mode counterpart to the ablation results in the main paper. The same trend remains under rollout generation: removing modality experts degrades audio fidelity and physical consistency, while removing stagewise training further weakens cross-modal alignment. In contrast, AV-CDiT achieves the best overall balance, especially on SSIM, ILD Error, AV-Lag Error, and AV-Corr Error, showing that both design choices remain important when predictions are recursively rolled out over time.

\begin{figure}[t]
\centering
\includegraphics[width=0.8\columnwidth]{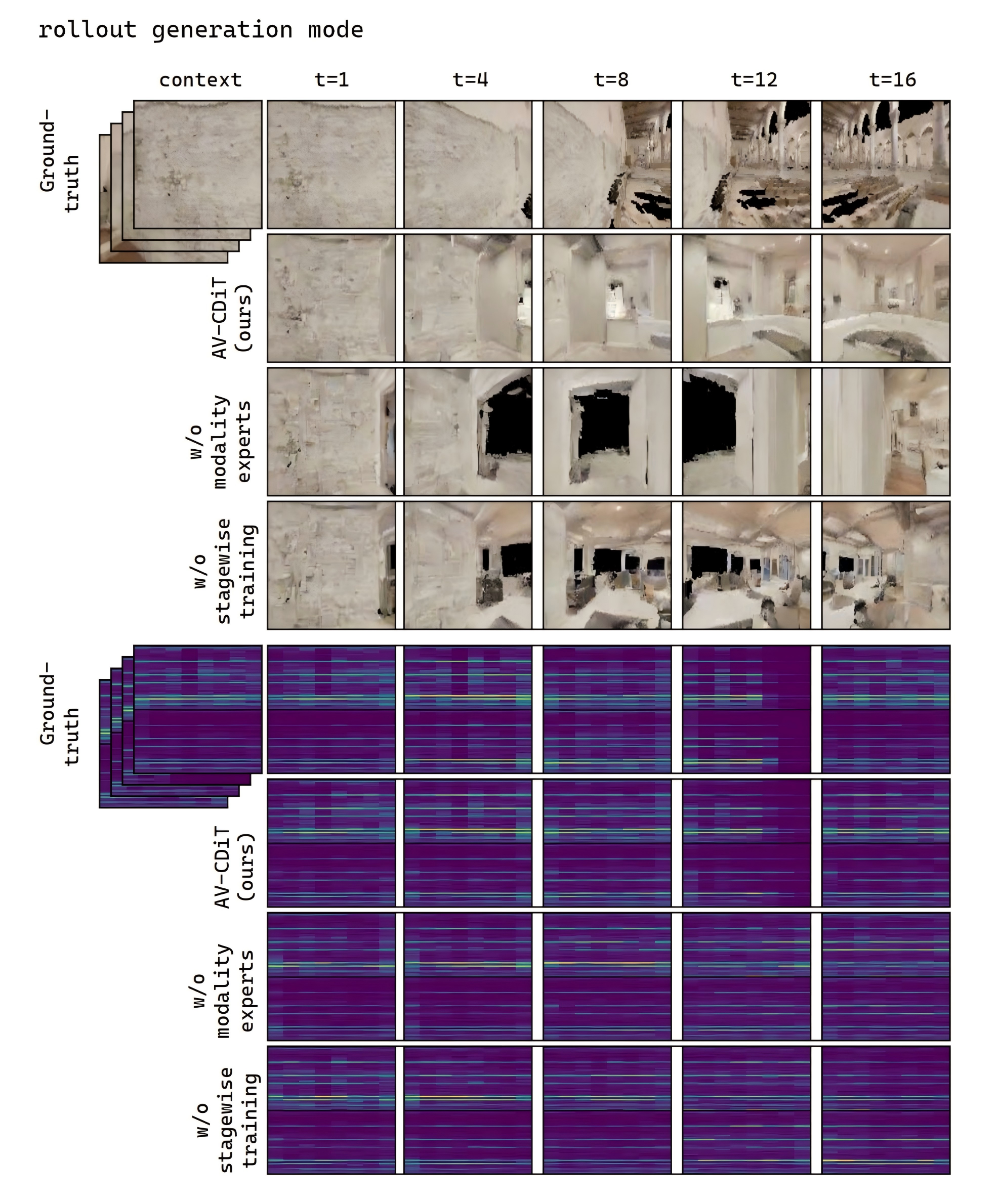}
\caption{Additional qualitative examples. This figure shows representative image and audio generation results of our model and two ablated variants under the rollout mode.}
   \label{fig:ablations2}
\end{figure}

Figure~\ref{fig:ablations2} presents qualitative comparisons among AV-CDiT and its ablated variants under rollout generation mode.

\begin{table*}[t]
  \centering
  \footnotesize
  \setlength{\tabcolsep}{3pt}
  \renewcommand{\arraystretch}{0.95}
  \begin{tabular}{@{}lcccccccc@{}}
  \toprule
  \textbf{Setting} &
  \multicolumn{3}{c}{\textbf{Vision Fidelity}} &
  \multicolumn{2}{c}{\textbf{Audio Fidelity}} &
  \multicolumn{3}{c}{\textbf{Physical Consistency}} \\
  \cmidrule(lr){2-4}
  \cmidrule(lr){5-6}
  \cmidrule(l){7-9}
  &
  \textbf{LPIPS}$\downarrow$ &
  \textbf{DreamSim}$\downarrow$ &
  \textbf{PSNR}$\uparrow$ &
  \textbf{LSD}$\downarrow$ &
  \textbf{SSIM}$\uparrow$ &
  \textbf{ILD}$\downarrow$ &
  \textbf{AV-Lag}$\downarrow$ &
  \textbf{AV-Corr}$\downarrow$ \\
  \midrule
  Mixed Sources & $0.381$ & $0.253$ & $16.459$ & $1.140$ & $0.431$ & $1.971$ & $1.240$ & $0.208$ \\
  Telephone Ring & $0.382$ & $0.255$ & $16.504$ & $1.311$ & $0.547$ & $2.471$ & $1.218$ & $0.194$ \\
  \bottomrule
  \end{tabular}
  \caption{Generalization to diverse sound-source categories under the \textbf{fixed-step} generation mode.}
  \label{tab:multisound-time}
  \end{table*}

  \begin{table*}[t]
  \centering
  \footnotesize
  \setlength{\tabcolsep}{3pt}
  \renewcommand{\arraystretch}{0.95}
  \begin{tabular}{@{}lcccccccc@{}}
  \toprule
  \textbf{Setting} &
  \multicolumn{3}{c}{\textbf{Vision Fidelity}} &
  \multicolumn{2}{c}{\textbf{Audio Fidelity}} &
  \multicolumn{3}{c}{\textbf{Physical Consistency}} \\
  \cmidrule(lr){2-4}
  \cmidrule(lr){5-6}
  \cmidrule(l){7-9}
  &
  \textbf{LPIPS}$\downarrow$ &
  \textbf{DreamSim}$\downarrow$ &
  \textbf{PSNR}$\uparrow$ &
  \textbf{LSD}$\downarrow$ &
  \textbf{SSIM}$\uparrow$ &
  \textbf{ILD}$\downarrow$ &
  \textbf{AV-Lag}$\downarrow$ &
  \textbf{AV-Corr}$\downarrow$ \\
  \midrule
  Mixed Sources & $0.406$ & $0.270$ & $15.881$ & $1.469$ & $0.533$ & $2.356$ & $1.557$ & $0.317$ \\
  Telephone Ring & $0.407$ & $0.272$ & $16.019$ & $1.620$ & $0.577$ & $2.654$ & $0.867$ & $0.322$ \\
  \bottomrule
  \end{tabular}
  \caption{Generalization to diverse sound-source categories under the \textbf{rollout} generation mode.}
  \label{tab:multisound-rollout}
  \end{table*}
  
\subsection{Source Source Diversity Generalizability}
\label{app:sound_source_diversity}

\noindent\textbf{Setup.}
To evaluate whether AV-CDiT generalizes to diverse sound-source categories rather than overfitting to a single target identity, we construct a generalized variant of AVW-4k by replacing the original telephone-ring source with a mixture of five common environmental sounds: door, fan, electric buzz, elevator, and engine.
The source category is randomly assigned while keeping the scene layouts, action space, and training pipeline unchanged, so the comparison isolates the effect of sound-source diversity.
We then train and test AV-CDiT on this mixed-source version of AVW-4k and compare it against the original telephone-ring setting under both the \textbf{fixed-step} and \textbf{rollout} generation modes.

\noindent\textbf{Results.}
As shown in Tables~\ref{tab:multisound-time} and~\ref{tab:multisound-rollout}, AV-CDiT maintains broadly stable performance on the multi-source benchmark.
Most visual metrics remain nearly unchanged compared with the original telephone-ring setting, indicating that increasing source diversity does not disrupt the model's visual dynamics prediction.
On the audio and cross-modal side, the mixed-source setting remains competitive and is favorable on LSD and ILD error in both generation modes, while the telephone-ring setting remains stronger on SSIM and AV-Lag.
The AV-Corr trend is mixed across the two settings but remains close overall.
Overall, these results suggest that AV-CDiT captures transferable audio-visual regularities rather than memorizing a single source type, demonstrating strong generalization across different environmental sound categories.

 \subsection{Reward Annotation and Prediction for Planning}
  \label{app:reward_annotation}

  For the planning experiments, we construct a reward-annotated version of AVW-4k using the reward formulation from continuous AV-Nav. The
  full navigation reward at timestep $t$ consists of a sparse success bonus, a constant step penalty, and a dense progress term:
  \begin{equation}
  R_t = r_{\mathrm{success}} + r_{\mathrm{step}} + r_{\mathrm{progress}}.
  \end{equation}
  The success bonus $r_{\mathrm{success}}$ is granted when the agent issues a \textit{stop} action within $1$m of the target sound source,
  while the step penalty $r_{\mathrm{step}}$ is applied at every timestep to encourage shorter trajectories.

  In our lookahead planner, AVWM is trained to predict the dense progress term,
  \begin{equation}
  r_{\mathrm{progress}} = d_{t-1} - d_t,
  \end{equation}
  where $d_t$ denotes the geodesic distance from the agent to the target sound source at timestep $t$. This reward is positive when the agent
  moves closer to the source and negative when it moves farther away. Compared with the sparse success signal, the progress term provides a
  denser action-level signal for ranking candidate rollouts.

  To enable reward prediction, we augment the diffusion target with an auxiliary reward token. The scalar reward is broadcast to the token
  dimensionality and concatenated with the visual and auditory latent tokens:
  \begin{equation}
  X_{t+\Delta t} = [h_{t+\Delta t}^v : h_{t+\Delta t}^a : h^r_{t+\Delta t}].
  \end{equation}
  The same diffusion objective is then applied to the augmented target sequence, allowing the model to jointly predict future visual
  observations, binaural audio, and the progress reward.

  During planning, the predicted progress reward is used only as an auxiliary scoring signal. For each candidate action sequence, the planner
  accumulates the discounted predicted progress rewards and combines them with the policy network's value estimate to obtain the branch score.
  The action sequence with the highest score is selected, and only its first action is executed in the real environment.

  We also evaluate reward prediction on held-out reward-annotated AVW-4k trajectories as a sanity check.
  We report reward prediction under two evaluation protocols.
  In fixed-step evaluation, the model predicts the cumulative progress reward over a future horizon $\Delta t$, and we report the resulting
  cumulative reward MSE.
  In rollout evaluation, the model is applied autoregressively and predicts the immediate one-step progress reward at each rollout step; we
  report the corresponding one-step reward MSE.
  These two metrics are therefore not directly comparable in magnitude, since they are defined on targets with different temporal
  aggregation and scale.
  The results are summarized in Table~\ref{tab:reward_mse}.

\begin{table}[t]
\centering
\scriptsize
\setlength{\tabcolsep}{3pt}
\renewcommand{\arraystretch}{0.95}
\begin{tabular}{lcc}
\toprule
  \textbf{Model} &
  \shortstack{\textbf{Fixed-step Cumulative}\\\textbf{Reward MSE}$\downarrow$} &
  \shortstack{\textbf{Rollout One-step}\\\textbf{Reward MSE}$\downarrow$} \\
\midrule
\shortstack{Reward-aware \\AV-CDiT} & $0.746$ & $0.052$ \\
\bottomrule
\end{tabular}
\caption{Held-out progress-reward prediction results for the reward-aware AV-CDiT used in planning.} %
\label{tab:reward_mse}
\end{table}

\newpage
\clearpage

\bibliography{aaai2027}

\end{document}